\documentstyle[prl,aps,epsfig,multicol]{revtex}
\begin{document}
\def\be{\begin{equation}}
\def\ee{\end{equation}}
\def\bearr{\begin{eqnarray}}
\def\eearr{\end{eqnarray}}
\def\tc{$T_c~$}
\def\tcl{$T_c^{1*}~$}
\def\c2{ CuO$_2~$}
\def\ruo{ RuO$_2~$}
\def\lsco{LSCO~}
\def\bi{bI-2201~}
\def\tl{Tl-2201~}
\def\hg{Hg-1201~}
\def\sro{$Sr_2 Ru O_4$~}
\def\rc{$RuSr_2Gd Cu_2 O_8$~}
\def\mgb{$MgB_2$~}
\def\pz{$p_z$~}
\def\ppi{$p\pi$~}
\def\sqo{$S(q,\omega)$~}
\def\tperp{$t_{\perp}$~}
\def\he4{${\rm {}^4He}$~}
\def\tss{${T_{ss}}$~}

\title{And quiet flows the supersolid \he4}

\author{ G. Baskaran \\
Institute of Mathematical Sciences\\
C.I.T. Campus,
Chennai 600 113, India }

\maketitle
\begin{abstract}
A superfluid having atomic scale superflow of a hexagonal lattice
of vortex and antivortex filaments, described by a single macroscopic
wave function is presented as a supersolid. As superfluid \he4
is pressurized, at a first order transition, rotons 
(atomic scale current circulation, a vortex loop) not only condense 
but also {\em expand and fuse into hexagonal or other 
complex superflow patterns}. The vortex 
core contains an excess density of non-condensate atoms. 
Further, a Kelvin (m = 0,
necklace) mode condenses in the vortex filaments. It results in a 3D
atom density wave of hcp symmetry. In our theory, superfluid
phase stiffness, rather than atom localization, imitates a solid
like rigidity.
\end{abstract}

\begin{multicols}{2}[]
Supersolid\cite{anlif,chester} is a quantum crystal that exhibits 
non classical moment of inertia\cite{leggett1}, 
a type of superfluid response.
Decades of efforts\cite{ExpReview}, following early theoretical 
suggestions, have 
culminated in a recent striking observation of non classical moment 
of inertia, by Kim and Chan\cite{kc1kc2}. This is 
yet another jewel in the crown of condensed \he4, an elusively
simple one component bose system; the deeper one searches and digs, 
the more surprises are in store.
This work has excited a renewed interest\cite{theory,pwaSS}
in the quantum many body 
theory of supersolid. It is likely to open new 
directions and find new phenomena in experiments; it also offers a good 
opportunity to bring the field of cold atom BEC, boson Mott insulators 
etc., closer to supersolid \he4. A correct theoretical understanding
of the supersolid phenomena will throw new light into old experimental 
quantum anomalies in the superfluid-solid helium 
interface\cite{RussianBook}, such as 
crystallization waves, Kapitza resistance etc.     

The supersolid predicted by Andreev, Lifshitz\cite{anlif} 
and Chester\cite{chester} has a 
reference crystal of localized atoms; large amplitude quantum
fluctuations create ground state vacancies or defects which 
undrgo condensation. In our mechanism superfluid is the
reference system and solid like rigidity 
emerges from the superfluid stiffness in a fundamental way.
A spatially periodic,
atom scale superfluid flow develops spontaneously in the ground 
state leading to a solid like response, in addition to 
superfluidity. We call it a roton fusion(ROFU) mechanism, as 
the atom scale flow pattern arises spontaneously
from a condensation and a complex fusion of real rotons (vortex
loop of atomic dimensions),
as we pressurize superfluid \he4. ROFU mechanism is one collective way of 
releasing the kinetic energy frustration, that is present in a solid with
localized atoms. It is a kind of spontaneous generation of 
staggered Abrikosov vortex lattice, created by 
hydrostatic pressure. Hydrostatic pressure, a scalar, mimics an 
internal staggered magnetic field.

Roton minimum of superfluid \he4 has been viewed\cite{enz}, 
for quite some  
time as a soft mode that drives a superfluid-solid phase 
transition, as rotons have wave vectors close to the 
reciprocal lattice vectors of the hcp \he4 crystal.
A new and crucial ingredient in our theory is to use the 
non-trivial inner structure of roton, an atomic size quantized
vortex loop. In our theory rotons do more than condensation; they 
expand and fuse into ordered vortex and antivortex filaments
(atom scale thickness)  and 
loose their identity. 
\begin{figure}[h]
\epsfxsize 8cm
\centerline{\epsfbox{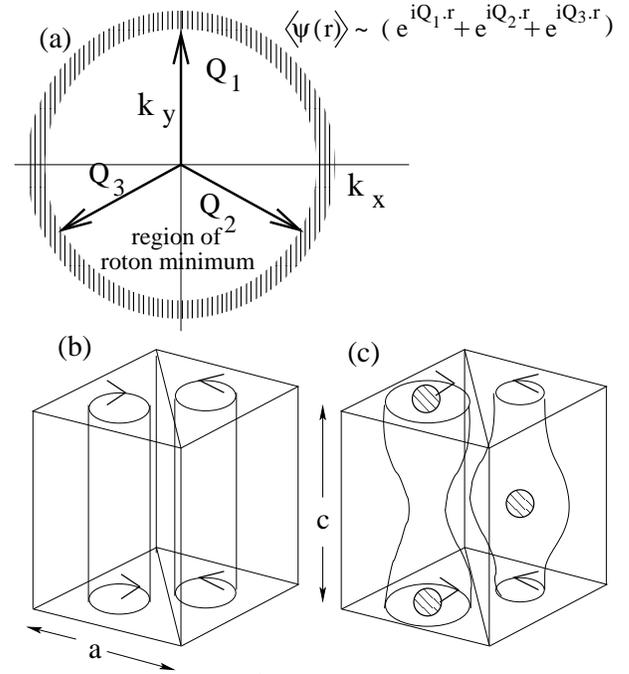}}
\caption{ Figure 1. a) Region of roton minima in the $(k_x,k_y,0)$
plane. ${\bf Q}_1$, ${\bf Q}_2$ and ${\bf Q}_3$ are
Fourier component vectors of the macroscopic 
condensate wave function $\langle \Psi({\bf r})\rangle$.
b) Unit cell of the hexagonal vortex-antivortex array (tubes); 
it has a 2D atom density wave of hexagonal
symmetry c) Kelvin (azimuthal quantum number
 m = 0, necklace) 
mode with a finite $\pm Q_z$ condenses in the vortex in a 
staggered fashion. It results in an atom density wave of hcp 
lattice. Lattice sites (not position of localized atoms) are 
denoted by shaded circle.} 
\end{figure}

In superfluid \he4, nearly 90 \% of the atoms are out side
the zero momentum condensate. These large fraction of atoms
with finite momenta appear as spatially uniform quantum 
fluctuation of the superfluid vacuum. They are part of the 
vacuum and the ground state is a 100 \% superfluid.
When vortices and antivortices are spontaneously created in
the ground state, a part of the non-condensate fraction gets 
piled up in the vortex core region, creating a net atom 
 density 
wave of hcp symmetry.  The self consistent potential 
 that maintains 
this redistribution arises from the underlying
 superfluid stiffness. 
In this sense superfluid stiffness
 gives a solid like rigidity to 
the 3D atom density wave.
 A fragile 2\% superfluid condensate(as 
measured by Kim and 
 Chan), by a complex maneuver of its own flow, 
orchestrates 
 the 3D atom density wave and a rich `lattice dynamics'.
Periodic superflow leads to a periodic modulation of quantum 
fluctuation and non condensate component.

In what follows we develop some heuristic pictures, followed by 
a Bogoliubov theory and then discuss some consequences. 

Kim and Chan have discovered non-classical moment of 
inertia in pressurized solid \he4; they observe strong 
thermal hysteresis, small variation of supersolid transition 
temperature \tss for a pressures range, 25 to 60 bars, and an 
anomalous increase of \tss with addition of ${\rm {}^3He}$ 
impurity. More importantly, the authors indicate that the 
superfluid response may not be due to zero point vacancies or 
defects or interfaces. That is, it is likely one is dealing 
with a solid with integer number of atoms per unit cell and 
yet exhibiting a superfluid property. 

While building our theory we remember that 
at the superfluid-solid first order transition at T = 0,
inter \he4 atom distance decreases only by a small, $\sim 3 \%$,
(a density decrease $\sim 10 \%$). This means that locally
the quantum solid is no more crowded than the quantum liquid is. 
Thus any generalized rigidity that emerges on both sides is 
likely to have the same quantum character, namely some kind of
phase stiffness (ODLRO) arising from local coherent number 
fluctuations. Further, the superfluid
solid coexistence line in the P-T plane has nearly zero slope 
for $T < 0.2$ K indicating 
that the two phases have the same entropy per mole, according
to Clausius-Clapeyron equation $\frac{dP_m}{dT} =
\frac{S_L - S_S}{V_L - V_S}$. This means that density of 
states of low energy bosonic quasi particles is nearly 
the same for superfluid and supersolid at the coexistence 
line, indicating a possible deep connection between the
stiffness of a superfluid and rigidity of a supersolid.

Now we present a Bogoliubov theory to illustrate our ROFU mechanism. 
Bogoliubov theory is a mean field theory that works well for weakly 
interacting bosons. However, for a proper choice two body pseudo
potential, it captures qualitative features and some
quantitative features, including roton spectrum. The model 
Hamiltonian is:
\be
H = \sum 
(\epsilon_k -\mu)
b^\dagger_{\bf k}
b^{}_{\bf k} +
\frac{1}{\Omega}\sum V({\bf q}) b^\dagger_{\bf k}
b^{}_{\bf k-q} b^\dagger_{\bf k'}
b^{}_{\bf k' + q}
\ee
Here $\epsilon_k = \frac{\hbar^2}{2M}$ is the kinetic energy
of \he4 atoms with mass M, chemical potential $\mu$, created
by operators $b^{\dagger}$'s.  $V({\bf q})$ is the two body, 
effective or pseudo potential and $\Omega$ is the volume of the 
system.

In the uniform superfluid phase, Bogoliubov
theory starts by replacing the zero momentum
operator $b^{}_0 \rightarrow n_0^{\frac{1}{2}}e^{i\phi}
N^{\frac{1}{2}}$, a 
classical expectation value, with $n_0$ as a condensate 
fraction.
The Bogoliubov quasi particle spectrum of the uniform 
superfluid state has the well known form 
\be
\hbar\omega_k = \sqrt{\left(\frac{\hbar^2k^2}{2M}\right)^2 + 
\left( \frac{\hbar^2 k^2}{2M}\right)  2 n_0 V(k)}
\ee
where $n_0$ is the zero momentum condensate fraction. When 
the potential $V(k)$ is negative for a range of k, the spectrum
has a roton minimum; an example is Bruckner
pseudo potential, $V(k) \approx V_0\frac{\sin ka_0}{ka_0}$. 
with $a_0~ \approx ~ 2.1 \mathring{\rm A}$ and 
$V_0 \approx 15~K$.

Landau suggested roton as the first excitation that has associated 
with it a rotational velocity flow (and hence the name roton). 
Feynman viewed it as a single atom motion, but dressed by a 
back flow, the net effect being a quantized vortex loop of atomic
dimensions.
Recent variational study and wave packet analysis\cite{rotonReatto} 
confirms this and shows roton wave packet as a ball of
current disturbance, and little density variation. 

Within the Bogoliubov theory, the roton minimum (figure 1a)
reaches zero energy (become completely soft), when
$V(k) = - \frac{1}{2n_0}\frac{\hbar^2 k^2}{2M}$.
This indicates an instability of 
the uniform superfluid solution. We expect a roton condensation 
and reorganization of the ground state. 
A complex current flow associated with a roton, rather than a 
simple density variation, made us wonder what will be the 
consequence of a roton condensation. It became clear that 
{\em a primary consequence of roton condensation will be formation
of a current density wave, not atom density wave}.
This resulted in our roton fusion hypothesis. 
According to this hypothesis the 
condensed rotons expand and fuse into a hexagonal array of
vortices and antivortices. Our hypothesis is naturally 
influenced by the known hcp structure of solid \he4.

To test our hypothesis we have compared the energy of uniform
superfluid state with an off diagonal long range order (ODLRO) 
ansatz that contains a hexagonal array of vortices and 
antivortices. After some struggle we found a simple and elegant
ansatz - in phase superposition of three plane waves generates 
the desired vortex structure:
\bearr
\langle \Psi({\bf r})\rangle  \equiv \psi_0({\bf r})\sim
n_0^{\frac{1}{2}}e^{i\phi}(
e^{i{\bf Q}_1 \cdot {\bf r}}+
e^{i{\bf Q}_2 \cdot {\bf r}}
+ e^{i{\bf Q}_3 \cdot {\bf r}})
\eearr
It also gives a lower energy 
than the uniform superfluid state, for a choice of $V(k)$.
Here ${\bf Q}_1= \frac{2\pi}{a}(\frac{2}{\sqrt 3},0)$, 
${\bf Q}_{2,3} = \frac{2\pi}{a}(\frac{-1}{\sqrt 3},\pm 1)$,
${\bf Q}_1 + {\bf Q}_2 + {\bf Q}_3 = 0$, `a' is the lattice parameter
of the hexagonal lattice. Here, $n_0$ and `a'
are variational parameters. The above ansatz leads to variation
of atom density only in the xy-plane, a 2D atom density wave.
This results in a supersolid behaviour along the x and y
directions and superfluid in z-direction - it is an anisotropic 
supersolid, a type of quantum liquid crystal.   Later we 
will improve it to get a 3D atom density wave of hcp 
symmetry. We can find more complex flow pattern (braided 
and knotted vortices) leading to an
atom density wave of hcp symmetry using more ${\bf Q}$'s etc.

One of the triangular sub lattices of the hexagonal lattice contains
vortices and the other antivortices (figure 1b). It is easy to show 
that equation (3) has asymptotic form 
\be
\langle \Psi ({\bf r})\rangle
 \sim x+iy~~~({\rm vortex}),~x-iy~~~({\rm antivortex})
\ee
respectively around the vortex and antivortex filaments of the two
sub lattices. 

In addition to global U(1) symmetry (overall phase of  
$\langle \Psi({\bf r})\rangle $) and translational symmetry 
there is a discrete symmetry breaking arising from P\&T violation. 
That is, another degenerate solution,
not connected by a global phase rotation of equation (5) is
obtained by the replacement ${\bf Q}_i \rightarrow - {\bf Q}_i$
in equation (3); equivalently, by an interchange
of vortices and antivortices. 

According to equation (4) the condensate fraction vanishes quadratically 
as we approach the core of line vortices. Further, the fraction
of the total particles condensed in our macroscopic wave function 
is $ < 1 $. Where are the rest of non-condensed particles ?
In the Bogoliubov theory of uniform superfluid, these are the 
particles that carry non zero momenta and they appear as a spatially 
uniform quantum fluctuations of the zero momentum condensate. 
They influence, for example the superfluid stiffness, through 
the spectrum; otherwise these finite momentum components are
not visible in the low energy dynamics of the irrotational 
superfluid.

In our ROFU, a spatial density
variation of the non-condensate fraction is induced.
In particular they appear as extra atom density in the normal
core region of our ordered vortices, where the condensate density
vanishes. This is somewhat counter intuitive, as one expects a 
depletion of fluid density at the vortex core.
It is known from an early work of Fetter\cite{fetter}
that within Bogoliubov 
theory, in the vortex core region the local atom density increases 
compared to the uniform background atom density. This is a 
remarkable non-local quantum effect, appearing within the 
Bogoliubov theory, as explained by Fetter.  

We elaborate this point further, by comparing a reference
Bose Einstein condensed (BEC) state
\be 
\Psi_{\rm BEC} \sim \prod_i 
(e^{i{\bf Q}_1 \cdot {\bf r}}+
e^{i{\bf Q}_2 \cdot {\bf r}}
+ e^{i{\bf Q}_3 \cdot {\bf r}})
\ee
with the N-particle projected wave function of 
Bogoliubov theory for our ROFU solution:
\be
\Psi_{\rm N}[{\bf r}_i]
= \sum_{P} 
\chi ({\bf r}_{P1},{\bf  r}_{P2} ) \chi ({\bf r}_{P3}, {\bf r}_{P4}) .....
\chi ({\bf r}_{PN-1}, {\bf r}_{PN})
\ee
Summation over permutation P symmetrises the wave function.
The pair function $\chi({\bf r}_1, {\bf r}_2) \equiv
\eta ({\bf r}_1,{\bf r}_2) 
\psi_0 \left[ \frac{{\bf r}_1 + {\bf r}_2}{2} \right] $.
When the pair function $\eta({\bf r}_1, {\bf r}_2) = 1$, equation (6)
becomes the same as BEC (equation 5). In the BEC state, atom density, 
by construction, vanishes as we approach the core of the 
line vortices. In the Bogoliubov wave function, equation (6), the
pair function $\eta$ makes an important difference and creates a 
a pile up of atom density through pushing some of the non-condensate 
fraction to core regions.  The pair function $\eta$ represents, a 
repulsive correlation induced between any two particles in the
Bogoliubov theory. It shows how pairs of particle are pushed in 
and out of the condensate by the Bogoliubov process 
$ \langle b_{{\bf Q}_1}\rangle \langle b_{{\bf Q}_2}\rangle 
b^{\dagger}_{\bf k}b^{\dagger}_{{- \bf k} + {\bf Q}_1 +
{\bf Q}_2}$ etc., in a space dependent fashion. 

We want to make certain remarks about the nature of ODLRO
in our supersolid phase. Our primary order
is a long range order in momentum space, an ODLRO. Bosons 
condense in a single macroscopic wave function 
$\psi_0({\bf r})$ given by 
equation (5).It  has three
Fourier coefficients which are not independent: 
$\langle b_{\bf Q_1}\rangle $= 
$\langle b_{\bf Q_2}\rangle $ =  
$\langle b_{\bf Q_3}\rangle$ = $n_0^{\frac{1}{2}}e^{i\phi}$:
they have same amplitude and phase.

The spatial variation of atom density implied by our ODLRO
should not be thought of as a diagonal long range order (DLRO).
A spatial periodic ordering of a small fraction of the 
non-condensate fraction is induced self consistently in our theory.
For example in our Bogoliubov factorization we get an anomalous
term such as 
$ \langle 
b^{\dagger}_{\bf Q_1}\rangle
\langle b^{}_{\bf Q_2}\rangle
b^{\dagger}_{\bf k} 
b^{}_{\bf k - Q_1 + Q_2 }$.
A density wave of wave vector $({\bf Q_1 - Q_2})$ induced by
the above term has its origin in single particle condensation.
In principle we could generate such an anomalous term through
non-vanishing averages such as, $\sum_k\langle
b^{\dagger}_{\bf k} b^{}_{\bf k + Q_1 - Q_2}\rangle$.
This will be an independent diagonal long range order (DLRO)
parameter. 

The solution we have discussed so far, equation (3), has no
density variation along z-axis. In real \he4
we expect maximum amount of atom density to be concentrated
in the normal core region. Thus it has its natural tendency for 
local spatial order, arising from short range repulsions. 
We view this 
as a Kelvin (azimuthal quantum number m = 0, necklace or
sausage ) mode condenses (figure 1c) at wave vectors
$Q_z = \pm \frac{2\pi}{c}$ in the vortex filaments. We modify 
equation (3) and generate a periodic 
modulation of the vortex core size, along the vortex filament
in a staggered fashion,
\bearr
\langle \Psi ({\bf r})\rangle  &\sim& (x + iy) (1 + \epsilon \cos Q_zz)
  ~~{\rm and}~~ \nonumber \\ 
&\sim& (x-iy)(1 - \epsilon \cos Q_zz) 
\eearr
close to core of the vortices of the two sub lattices. In view of
the short range interaction of the model Hamiltonian,
the total energy is reduced further for small $\epsilon$, another 
variational parameter.  This vortex core modulation leads to an 
atom density wave of hcp symmetry and a 3D solid like rigidity. 

We can choose our model parameters of Hamiltonian (equation 1)
such that our hexagonal lattice contains an average of 2 atoms
per unit cell or close to it.

From our solution it follows that the superfluid density is
mostly concentrated outside the vortex core, which is
the interstitial region of the hcp atom density wave (figure 1b and
1c).

So far we have sketched a Bogoliubov theory, a ROFU solution and 
some key features of the ground state. Our theory is far from
rigorous and complete. However, the physical argument for a 
spontaneous generation of microscopic circulating ground state
current is compelling. Bogoliubov (mean field) theory, in view of 
its non perturbative character is capable of finding possible 
new phases in dense liquid \he4.

In what follows we show how our
theory qualitatively explains salient features of Kim-Chan's
results. Further, we briefly sketch interesting consequences 
that also follow; some of them are very unique to our theory. 

In our ROFU solution we 
discussed a 2D solid and a 3D solid. It is likely that as 
a function of pressure there is a small region in the 
pressure-temperature diagram where the 2D solid intervenes, 
as shown in figure 2.
\begin{figure}[h]
\epsfxsize 6.5cm
\centerline {\epsfbox{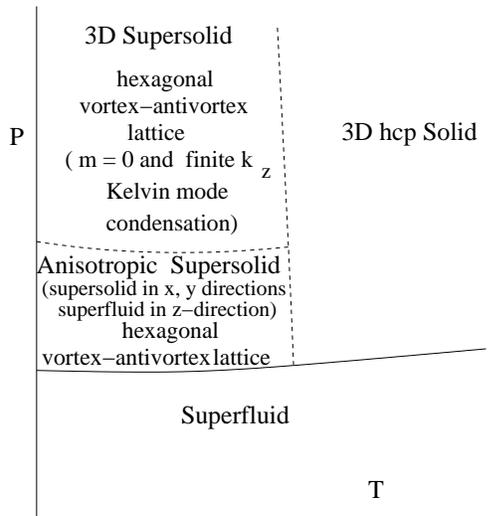}}
\caption{ Schematic phase diagram for \he4 in the pressure
temperature plane. An intermediate region of anisotropic
supersolid (supersolid in x, y direction and superfluid
in z-direction) is suggested.}
\end{figure}

We can qualitatively discuss some features of finite 
temperature phase transitions that are unique to ROFU solution.
Thermally produced vortices in the ROFU condensate will 
proliferate, depin and melt the underlying vortex lattice.
Disappearance of ODLRO will result in a solid of localized atoms. 
As the process is a melting of a vortex lattice, 
i) phase transition will be a first order one and ii) small 
traces of 
 ${}^3{\rm He}$ atoms could help pin the vortices 
and thereby increase the supersolid-solid transition temperature. 
Both are consistent with Kim-Chan's observations. The anomalously
low critical velocity, for destruction of the non classical 
moment of inertia (super solidity), observed by Kim and Chan
is likely to arise from a collective depinning of the ground
state vortices from their weak self consistent potential. 

A key prediction of our theory is the presence of atom scale 
circulation. One important consequence of this is on lattice 
dynamics. The P \& T violation in the ground state and the first 
order dynamics of vortices makes the lattice dynamics different 
from the classical hcp solid. We get splitting of degeneracies etc.; 
some of the anomalous modes can be viewed as coupled Kelvin modes. 
We will discuss them in a future publication. 

In summary, our supersolid is a superfluid in disguise with ineresting
consequences.  

It is a pleasure to thank N. Kumar for an early (1971) and 
insightful introduction to Leggett's paper\cite{leggett1} 
and a recent discussion; Dung-Hai Lee for alerting me of Kim and 
Chan's work\cite{kc1kc2}; Moses Chan for a brief discussion 
at Hong Kong, implying `no room for vacancy' superfluidity
in their results.

\end{multicols}

\begin{references}

\bibitem{anlif}A. F. Andreev, I.M. Lifshitz,
Sov. Phys. JETP, {\bf 29} 1107 (1969)
\bibitem{chester}G.V. Chester, Phys. Rev. {\bf A 2} 256 (1970)
\bibitem{leggett1}A. Leggett, Phys. Rev. Lett., {\bf 25} 1543 (1970)
\bibitem{ExpReview} M. W. Meisel, Physica {\bf B 178} 121 (1992);
J. M. Goodkind, Phys. Rev. Lett., {\bf 89} 095301 (2002)
\bibitem{kc1kc2} E. Kim and M.H.W. Chan, Nature {\bf 427}
225 (2004); Science, {\bf 305} 1941 (2004)
\bibitem{theory}D.M. Ceperley and B. Bernu,
Phys. Rev. Lett., {\bf 95}~155303  (2005); D. Weiss,
cond-mat/0407043; 
W. M. Saslow, cond-mat/0407166;
S. Koh, cond-mat/0411713;
N. Prokof'ev and B. Svistunov, cond-mat/0409472;
E. Burovski et al., cond-mat/0412644;
M. Tiwari and A. Datta, cond-mat/0406124
H. Zhai and Y.S. Wu, cond-mat/0501309;
X. Dai, M. Ma and F.C. Zhang, cond-mat/0501373
\bibitem{pwaSS}P.W. Anderson,
`Basic Notions of Condensed Matter Physics',
Ch 4, pp143, (Benjamin, 1984);
P. W. Anderson, cond-mat/0504731
\bibitem{RussianBook} A. Ya. Parshin, in Low temperature
Physics, Ed. A.S. Borovik-Romanov (MIR Publishers, Moscow, 
1985) pp.15; K. N. Zinov'ea, ibid. pp 78
\bibitem{enz}T. Schneider and C.P. Enz, Phys. Rev. Lett.,
{\bf 27} 1186 (1971)
\bibitem{rotonReatto}D.E. Galli, E. Cechetti and L. Reattto,
Phys. Rev. Lett., {\bf 77} 5401 (1996)
\bibitem{fetter}A. L. Fetter., Annals of Physics, {\bf 70} 67 (1972)
\end{references}
\end{document}